\documentclass[12pt]{article}
\usepackage{graphicx}

\newcommand\pubnumber{}
\newcommand\pubdate{\today}

\def\Title#1{\begin{center} {\Large #1 } \end{center}}
\def\Author#1{\begin{center}{ \sc #1} \end{center}}
\def\Address#1{\begin{center}{ \it #1} \end{center}}

\newcommand\pubblock{\rightline{\begin{tabular}{l} \pubnumber\\
         \pubdate  \end{tabular}}}
\newenvironment{Abstract}{\begin{quotation}  }{\end{quotation}}
\newenvironment{Presented}{\begin{quotation} \begin{center} 
             PRESENTED AT\end{center}\bigskip 
      \begin{center}\begin{large}}{\end{large}\end{center} \end{quotation}}

\begin{document}
\begin{titlepage}
\pubblock

\vfill
\Title{Theoretical and Experimental Status of Inclusive\\ Semileptonic
Decays and
 Fits for $|V_{cb}|$}
\vfill
\Author{Paolo Gambino$^{a}$ and Christoph Schwanda$^b$}
\Address{$^a$ Dipartimento di Fisica Teorica, Universit\`a di Torino\\
and INFN, sez. Torino, Via Giuria 1, 10125 Torino, ITALY}

\Address{$^b$ Institute of High Energy Physics, Austrian Academy of Sciences\\
Nikolsdorfer Gasse 18, 1050 Wien, AUSTRIA}
\vfill
\begin{Abstract}
We review recent experimental and theoretical developments  
in inclusive semileptonic  $B\to X_c\ell\nu$ decays. In particular, we discuss  the determination of $|V_{cb}|$ and of the heavy quark masses through 
fits based on 
the Operator Product Expansion.
\end{Abstract}
\vfill
\begin{Presented}
the 6$^{th}$ International Workshop\\ on the CKM
Unitarity Triangle (CKM2010)\\
University of Warwick, UK, 6--10 September 2010
\end{Presented}
\vfill
\end{titlepage}

Semileptonic $b\to c$ transitions  play an important role in various
aspects of heavy flavor physics. They allow for a determination of the
CKM matrix element $V_{cb}$, which  is a crucial input in all
Unitarity Triangle analyses: for instance the $(\bar\rho, \bar \eta)$
constraint determined by $\varepsilon_K$ is very sensitive to the
precise value of the Wolfenstein parameter $A$, which is essentially
determined by $|V_{cb}|$. Moreover, the bottom quark mass and the
hadronic parameters extracted from fits to inclusive semileptonic and
radiative moments are key inputs for the inclusive $|V_{ub}|$
determination, the normalization of rare $B$ decays like $B\to X_s
\gamma$, and various other $B$ physics applications.

We will review the present theoretical and experimental status of
inclusive $B\to X_c \ell\nu$ decays, with particular emphasis on
recent developments, among which new calculations and measurements.


\newcommand{\as}{\alpha_s}
\newcommand{\mupi}{\mu_\pi^2}
\newcommand{\mug}{\mu_G^2}
\newcommand{\rd}{\rho_D^3}
\newcommand{\rls}{\rho_{LS}^3}

\section{Theoretical framework}
Our understanding of inclusive semileptonic $B$ decays rests on a
simple idea: as all final states are summed over in inclusive decays, 
the final quark hadronizes with unit probability and the 
transition amplitude is sensitive only to the long-distance dynamics  
of the initial $B$ meson, which can indeed be factorized. An Operator
Product Expansion (OPE)  allows us to express the non-perturbative
physics in terms 
of matrix elements of local operators of dimension $d\ge 5$, while 
the Wilson coefficients are perturbative~\cite{Bigi:1992su,Blok:1993va}.
The leading term in this double expansion in  $\alpha_s$ and
$\Lambda_{\rm QCD}/m_b$  is given by the free $b$ quark decay, and
the first corrections are $O(\alpha_s)$ and $O(\Lambda^2_{\rm QCD}/m_b^2)$.
The relevant parameters are the heavy quark masses $m_b$ and  $m_c$,
the strong coupling $\alpha_s$, and
the matrix elements of the local operators: $\mupi$ and $\mug$ at
$O(1/m_b^2)$, $\rd$ and
 $\rls$ at $O(1/m_b^3)$, etc. Since the OPE is valid only for
sufficiently inclusive measurements and away
from perturbative singularities, the relevant quantities to be
measured are global shape
parameters (the first few moments of various kinematic distributions)
and the total rate. The former
give information on the masses and matrix elements, the latter on $|V_{cb}|$.
The OPE parameters describe universal properties of the $B$
meson and of the quarks and are useful in many applications.

The main ingredients for an accurate analysis of the experimental data
have been known for some time. Two implementations are currently
employed, based on either the kinetic
scheme~\cite{Benson:2003kp,Gambino:2004qm,Benson:2004sg} or the $1S$
scheme~\cite{Bauer:2004ve}. They both include terms through
$O(\alpha_s^2 \beta_0)$~\cite{pertb} and $O(1/m_b^3)$~\cite{nonpert}
but they use different perturbative schemes, include a somewhat
different choice of experimental data under specific assumptions, and
estimate the theoretical uncertainty in two distinct
ways. Nevertheless, it is reassuring that, as we will show below, the
two methods yield very close results for $|V_{cb}|$.

The complete two-loop perturbative corrections to the width and
moments of the lepton energy and hadronic mass distributions have been
recently computed 
\cite{czarnecki-pak,melnikov} by both numerical and analytic
methods. In general, using $\alpha_s(m_b)$ in the on-shell scheme, the
non-BLM corrections amount to about $-20\%$ of the two-loop BLM
corrections. In the kinetic scheme with cutoff $\mu=1$GeV, the
perturbative expansion of the total width is
\begin{equation}
  \Gamma[\bar{B} \to X_c e \bar{\nu}] \propto 1 - 0.96\, \frac{\as(m_b)}{\pi} -0.48 \beta_0 \left( \frac{\as}{\pi} \right)^2 + 0.82 \left( \frac{\as}{\pi} \right)^2 + O(\as^3) \approx 0.916
  \label{expkin}
\end{equation}
Higher order BLM corrections to the width and moments are also
known~\cite{Benson:2003kp,pertb}. The resummed BLM result is
numerically
very close to the NNLO one~\cite{Benson:2003kp}. The residual
perturbative error in the total width is therefore about 1\%.

Since the numerical results of \cite{melnikov} are available for a
variety of lepton energy cuts and values of $m_c/m_b$, it is now
possible to implement them in a global fit.
In the normalized leptonic moments the perturbative  corrections
cancel to large extent, independently of the scheme, as hard gluon
emission is comparatively suppressed. This
pattern of cancellations, crucial for a correct estimate of the
theoretical error, is confirmed by the complete $O(\alpha_s^2)$
calculation, although the numerical precision of the available results
is not always sufficient to improve the final accuracy. The actual
implementation in the kinetic scheme is under way.

Another source of significant theoretical uncertainty are the $O(\as
\Lambda^2_{QCD}/m_b^2)$ corrections to the width and to the
moments. Only the $O(\as \mu^2_{\pi}/m_b^2)$ terms are
known~\cite{Becher:2007tk}. A complete calculation of these effects
has been recently performed in the case of inclusive radiative
decays~\cite{ewerth}, where the $O(\as)$ correction increase the
coefficient of $\mug$ in the rate by almost 20\%. The extension of this
calculation to the semileptonic case is in progress. In view of the
numerical importance of $O(1/m_b^3)$ corrections, if the 1\% precision
in the width is to be reached, the effects $O(\as/m_b^3)$ will also be
necessary.

For what concerns the higher order power corrections, a thorough
analysis of $O(1/m_b^4)$ and $O(1/m_Q^5)$ effects has just
appeared~\cite{Mannel:2010wj}. The main problem is the proliferation
of non-perturbative  parameters: {\it e.g.}\ as many as nine new
expectation values appear at $O(1/m_b^4)$. As they cannot be fitted
from experiment, in Ref.~\cite{Mannel:2010wj} they are estimated in
the ground state saturation approximation, reducing them to the known
$O(1/m_b^{2,3})$ parameters. In this approximation the total
$O(1/m_Q^{4,5})$ correction to the width is about +1.3\%. The
$O(1/m_Q^5)$ effects are dominated by $O(1/m_b^{3} m_c^2)$ Intrinsic
Charm contributions, amounting to +0.7\%~\cite{Bigi:2009ym}. The
actual effect on $|V_{cb}|$ depends also on the corrections to the
moments. The authors of \cite{Mannel:2010wj} estimate that the overall
effect on $|V_{cb}|$ is a 0.4\% increase, consistent with our
preliminary implementation of these effects in the kinetic global
fit. While this sets the scale of  higher order power corrections, it
is yet unclear how much the result depends on the assumptions made on
the expectation values.

The first two moments of the photon energy distribution in $B\to X_s
\gamma$ are also routinely included in the semileptonic fit. They are
sensitive to $m_b$ and $\mu_\pi^2$ in particular. However, the
experimental lower cut on the photon energy introduces a sensitivity
to the Fermi motion of the $b$-quark inside the $B$ meson and tend to
disrupt the OPE. One can still resum the higher-order terms into a
non-local distribution function and since the lowest integer moments
of this function are given in terms of the local OPE parameters, one
can parameterize it assuming different functional
forms~\cite{Benson:2004sg}. Another serious problem is that only the
leading operator contributing to inclusive radiative decays admits an
OPE. Therefore  in principle unknown $O(\alpha_s\Lambda/m_b)$
contributions should be expected~\cite{paz} and radiative moments
should be considered with care in the context of high precision
analyses.

\section{Measurements of moments}

BaBar has recently published a study of the hadronic mass spectrum
$m_X$ in inclusive decays $B\to
X_c\ell\nu$~\cite{Aubert:2009qda}. The main steps of this analysis, based on a
data sample of 232 million $\Upsilon(4S)\to B\bar B$~events, are:
First, the decay of one $B$~meson in the event is fully reconstructed
in a hadronic mode ($B_\mathrm{tag}$) and the associated tracks and
clusters are removed from the event. Such a
$B_\mathrm{tag}$~candidate can be found in about 0.4\% of the
$\Upsilon(4S)$~events with a signal purity of about 80\%. Then, the
semileptonic decay of the second $B$~meson in the event
($B_\mathrm{sig}$) is selected by searching for an identified
charged lepton (electron or muon) with momentum above
0.8~GeV/$c$. Finally, all remaining particles in the event are
combined to reconstruct the hadronic $X$~system. The resolution in
$m_X$ is improved by a kinematic fit taking into account 4-momentum
conservation and the consistency of the missing mass with a zero mass
neutrino.

Still, the observed $m_X$~spectrum is distorted by resolution and
acceptance effects and cannot be used directly to obtain the hadronic
mass moments. BaBar implements a linear correction to obtain the true
moments from the reconstructed ones. Different corrections are applied
depending on the $X$~system multiplicity,
$E_\mathrm{miss}-cp_\mathrm{miss}$ and the lepton momentum. In this
way, BaBar measures the moments of the hadronic mass spectrum up to
$\langle m^6_X\rangle$ for minimum lepton energies ranging between 0.8
and 1.9~GeV.

This study also updates the previous BaBar measurement of the lepton
energy moments in $B\to X_c\ell\nu$~\cite{Aubert:2004td} using
new branching fraction measurements for background decays and
improving the evaluation of systematic uncertainties. Also, the first
measurement of combined hadronic mass and energy moments $\langle
n^k_X\rangle$ with $k=2,4,6$ is presented, where the latter are
defined as $n^2_X=m^2_Xc^4-2\tilde\Lambda E_X+\tilde\Lambda^2$,
with $m_X$ and $E_X$ the mass and the energy of the $X$~system and
$\tilde\Lambda$ a constant fixed to 0.65~GeV.

BaBar interprets their data using the OPE  in the kinetic
scheme~\cite{Benson:2003kp,Gambino:2004qm,Benson:2004sg} and performs
a simultaneous fit to 12 hadronic mass moments (or 12 combined
mass-energy moments), 13 lepton energy moments (including partial
branching fractions as 'zero order' moments), and 3 photon energy
moments in $B\to X_s\gamma$~\cite{Aubert:2005cua,Aubert:2006gg}. The
results are given in Table~\ref{tab:1}.
\begin{table}
  \caption{Results of the OPE fits in the kinetic scheme to the BaBar
    data~\cite{Aubert:2009qda}. The first uncertainty quoted is
    experimental, the second theoretical.} \label{tab:1}
  \begin{center} \begin{tabular}{c|cc}
    \hline \hline
    & Hadronic moments & Mass-energy moments\\
    \hline
    $|V_{cb}|$ (10$^{-3}$) & $42.05\pm 0.45\pm 0.70$ & $41.91\pm
    0.48\pm 0.70$\\
    $m_b$ (GeV) & $4.549\pm 0.031\pm 0.038$ & $4.556\pm
    0.034\pm 0.041$\\
    $\mathcal{B}(B\to X_c\ell\nu)$ (\%) & $10.64\pm 0.17\pm 0.06$ &
    $10.64\pm 0.17\pm 0.06$\\
    $\chi^2$/ndf. & 10.9/28 & 8.2/28\\
    \hline \hline
  \end{tabular} \end{center}
\end{table}

Also the Belle collaboration has obtained measurements of the lepton
energy~$E_\ell$ and the hadronic mass spectrum~$m_X$ in $B\to
X_c\ell\nu$ using 152~million $\Upsilon(4S)\to B\bar
B$~events~\cite{Urquijo:2006wd,Schwanda:2006nf}. The experimental
method is similar to the BaBar analysis discussed previously, {\it
  i.e.}, one $B$~meson is fully reconstructed in a hadronic mode and
an identified lepton is required to select semileptonic decays of the
second $B$. In the Belle analyses acceptance and finite resolution
effects in the $E_\ell$ and $m_X$~spectra are corrected by unfolding
using the SVD algorithm~\cite{Hocker:1995kb}. Belle measures $\langle
E^k_\ell\rangle$ for $k=0,1,2,3,4$ and minimum lepton energies ranging
between 0.4 and 2.0~GeV. Moments of the hadronic mass~$\langle
m^k_X\rangle$ are measured for $k=2,4$ and minimum lepton energies
between 0.7 and 1.9~GeV.

To obtain $|V_{cb}|$, Belle fits 14 moments of the lepton energy
spectrum, 7 hadronic mass moments and 4 moments of the photon energy
spectrum in $B\to X_s\gamma$~\cite{Schwanda:2008kw} to OPE expressions
derived in the
kinetic~\cite{Benson:2003kp,Gambino:2004qm,Benson:2004sg} and
1S~schemes~\cite{Bauer:2004ve}. Both theoretical frameworks are
considered independently and yield  very consistent results with the
Belle data, Table~\ref{tab:2}.
\begin{table}
  \caption{Results of the OPE fits in the kinetic and 1S schemes to
    the Belle data~\cite{Schwanda:2008kw}.} \label{tab:2}
  \begin{center} \begin{tabular}{c|cc}
    \hline \hline
    & Kinetic scheme & 1S scheme\\
    \hline
    $|V_{cb}|$ (10$^{-3}$) & $41.58\pm 0.90$ & $41.56\pm 0.68$\\
    $\mathcal{B}(B\to X_c\ell\nu)$ (\%) & $10.49\pm 0.23$ & $10.60\pm
    0.28$\\
    $\chi^2$/ndf. & 4.7/18 & 7.3/18\\
    \hline \hline
  \end{tabular} \end{center}
\end{table}

\section{Global HFAG fit}

The Heavy Flavor Averaging Group (HFAG) has performed as global
analysis of inclusive observables in $B\to X_c\ell\nu$ and $B\to
X_s\gamma$ decays to determine $|V_{cb}|$, the $b$-quark mass $m_b$
and the higher order parameters in the OPE description of these
decays. This analysis combines data from the BaBar, Belle, CLEO, CDF
and DELPHI experiments.

The global fit is done both with expressions derived in the
kinetic~\cite{Benson:2003kp,Gambino:2004qm,Benson:2004sg} and
1S~schemes~\cite{Bauer:2004ve}. In both cases 7 free parameters are
determined in the fit. The only external input used in the analysis is
the average $B^0$ and $B^+$~lifetime.

The data used in the global fit is listed in Table~\ref{tab:3}. In
total 66 measurements -- 29 from BaBar, 25 from Belle and 12 from
other experiments -- are used. Note that the analysis in the 1S scheme
still uses the BaBar 2004 hadronic moment measurements~\cite{Aubert:2004tea}.
\begin{table}
  \caption{Experimental data used in the HFAG analysis of inclusive
    $B\to X_c\ell\nu$ and $B\to X_s\gamma$ decays. In the table,
    $\langle E^k_\ell\rangle$, $\langle m^k_X\rangle$ and $\langle
    E^k_\gamma\rangle$ refer to the moments of the lepton energy and hadronic
    mass spectrum in $B\to X_c\ell\nu$ and to the photon energy
    moments in $B\to X_s\gamma$, respectively. The index $k$ specifies
    the order of the moments used.} \label{tab:3}
  \begin{center} \begin{tabular}{l|l}
    \hline \hline
    BaBar & $\langle E^k_\ell\rangle$:
    $k=0,1,2,3$~\cite{Aubert:2009qda,Aubert:2004td}, $\langle
    m^k_X\rangle$: $k=2,4,6$~\cite{Aubert:2009qda}, $\langle
    E^k_\gamma\rangle$: $k=1,2$~\cite{Aubert:2005cua,Aubert:2006gg}\\
    \hline
    Belle & $\langle E^k_\ell\rangle$:
    $k=0,1,2,3$~\cite{Urquijo:2006wd}, $\langle m^k_X\rangle$:
    $k=2,4$~\cite{Schwanda:2006nf}, $\langle E^k_\gamma\rangle$:
    $k=1,2$~\cite{:2009qg}\\
    \hline
    CDF & $\langle m^k_X\rangle$: $k=2,4$~\cite{Acosta:2005qh}\\
    \hline
    CLEO & $\langle m^k_X\rangle$: $k=2,4$~\cite{Csorna:2004kp},
    $\langle E^k_\gamma\rangle$: $k=1$~\cite{Chen:2001fja}\\
    \hline
    DELPHI &  $\langle E^k_\ell\rangle$:
    $k=1,2,3$~\cite{Abdallah:2005cx}, $\langle m^k_X\rangle$:
    $k=2,4$~\cite{Abdallah:2005cx}\\
    \hline \hline
  \end{tabular} \end{center}
\end{table}

\begin{figure}[t]
  \begin{center}
    \includegraphics[width=0.45\columnwidth]{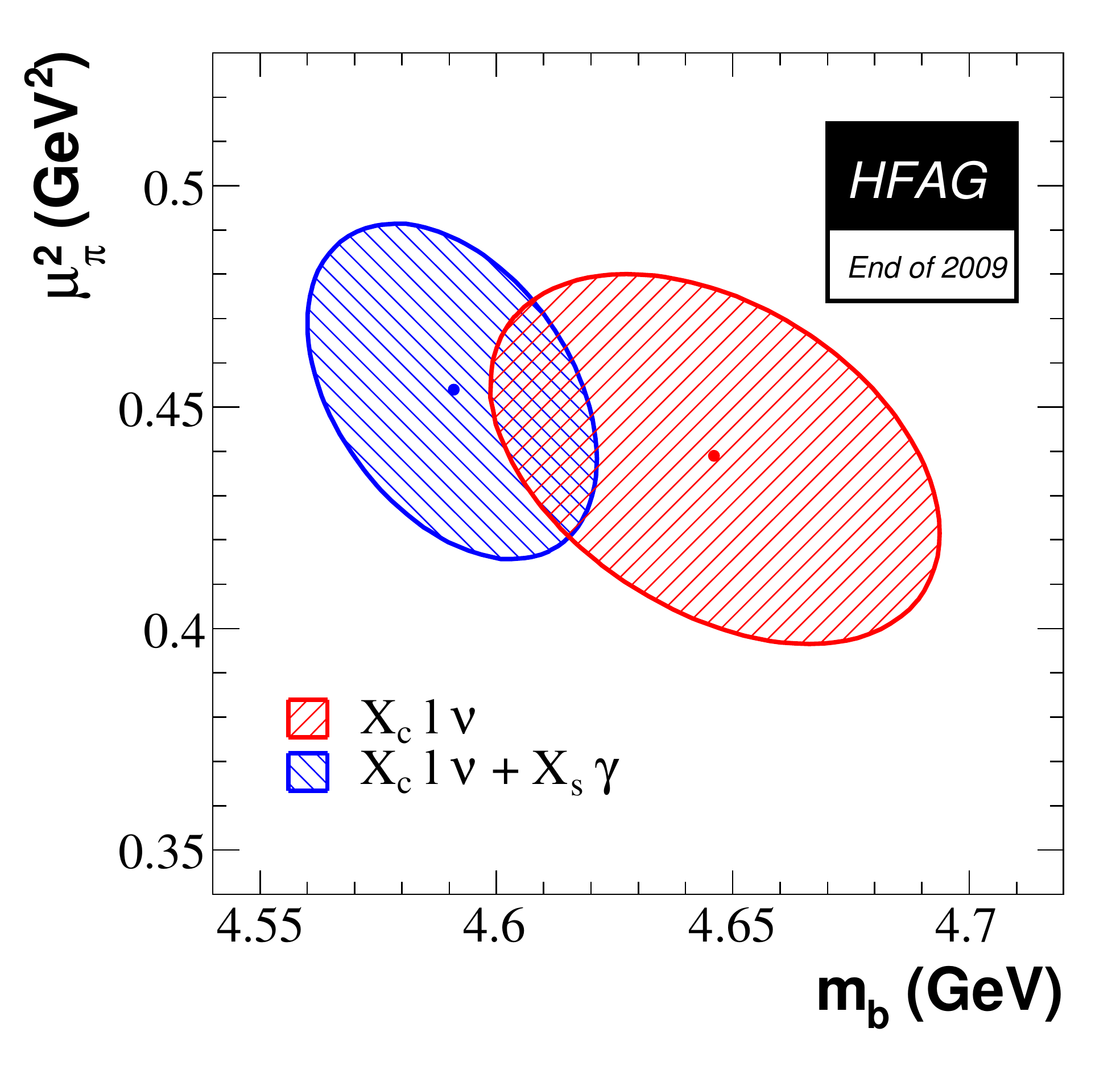}
    \includegraphics[width=0.45\columnwidth]{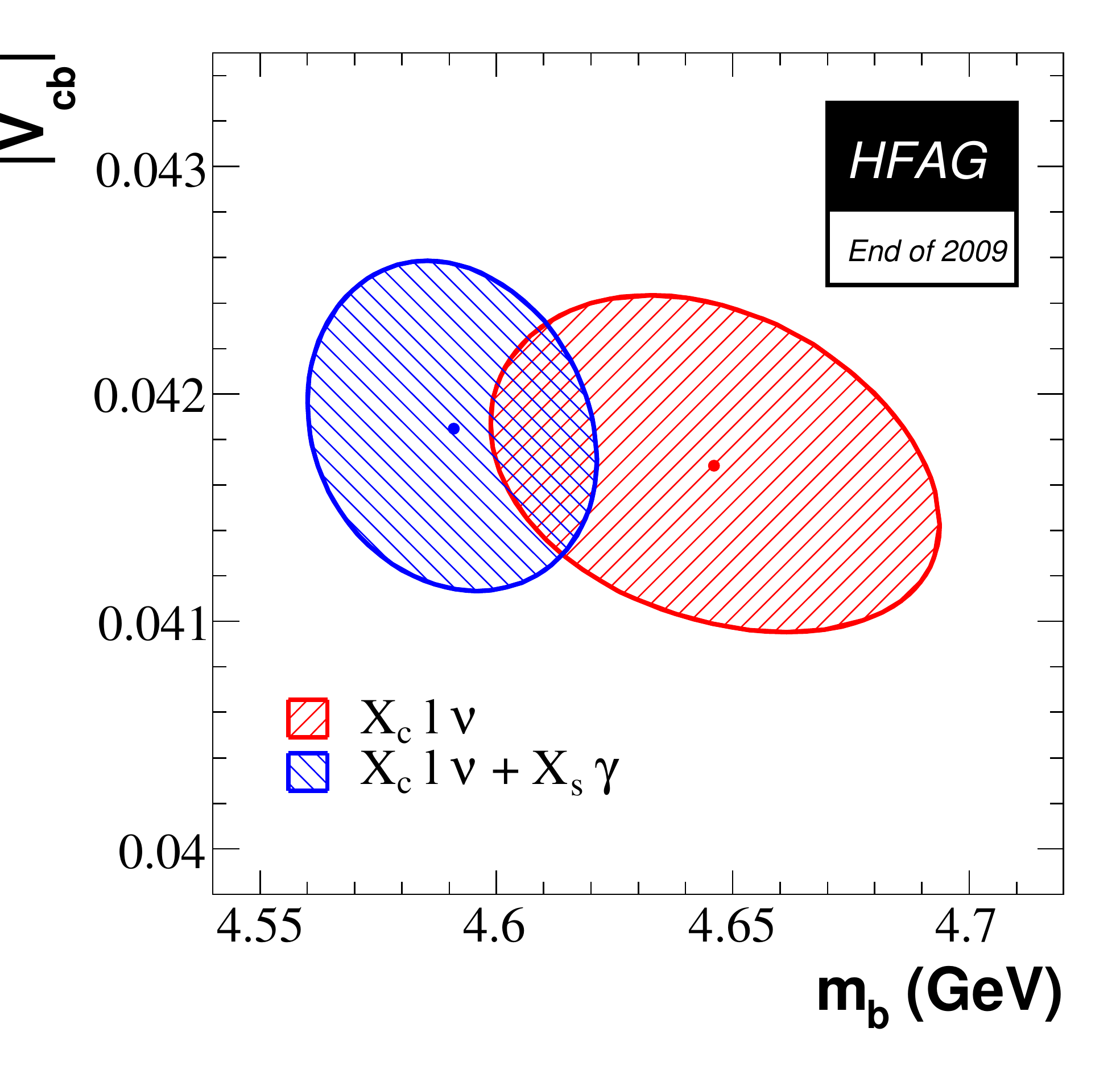}
  \end{center}
  \caption{$\Delta\chi^2=1$~contours for the HFAG global fit in the
    kinetic mass scheme.} \label{fig:1}
\end{figure}

The results of the global fit in the kinetic scheme are given in
Table~\ref{tab:4} and Fig.~\ref{fig:1}. The results of the 1S scheme
analysis are shown in Table~\ref{tab:5}. In both cases, the results
with all moments and with $B\to X_c\ell\nu$~moments only are quoted.
\begin{table}
  \caption{Results of the HFAG global fit in the kinetic scheme. The
  first error on $|V_{cb}|$ is the uncertainty from the global fit,
  the second is the error in the average $B$~lifetime and the third
  error is an additional theoretical uncertainty arising from the
  calculation of $|V_{cb}|$.} \label{tab:4}
  \begin{center} \begin{tabular}{c|cccc}
    \hline \hline
    Input & $|V_{cb}|$ (10$^{-3}$) & $m^\mathrm{kin}_b$ (GeV) &
    $\mu^2_\pi$ (GeV$^2$) & $\chi^2/$ndf.\\
    \hline
    all moments & $41.85\pm 0.42\pm 0.09\pm 0.59$ & $4.591\pm 0.031$ &
    $0.454\pm 0.038$ & 29.7/59\\
    $X_c\ell\nu$ only & $41.68\pm 0.44\pm 0.09\pm 0.58$ & $4.646\pm
    0.047$ & $0.439\pm 0.042$ & 24.2/48\\
    \hline \hline
  \end{tabular} \end{center}
\end{table}
\begin{table}
  \caption{Results of the HFAG global fit in the 1S scheme.}
 \label{tab:5}
  \begin{center} \begin{tabular}{c|cccc}
    \hline \hline
    Input & $|V_{cb}|$ (10$^{-3}$) & $m^{1S}_b$ (GeV) &
    $\lambda_1$ (GeV$^2$) & $\chi^2/$ndf.\\
    \hline
    all moments & $41.87\pm 0.25$ & $4.685\pm 0.029$ & $-0.373\pm
    0.052$ & 32.0/57\\
    $X_c\ell\nu$ only & $42.31\pm 0.36$ & $4.619\pm 0.047$ &
    $-0.427\pm 0.057$ & 24.2/46\\
    \hline \hline
  \end{tabular} \end{center}
\end{table}

\section{Discussion}

\begin{figure}
  \begin{center}
    \includegraphics[width=0.515\columnwidth]{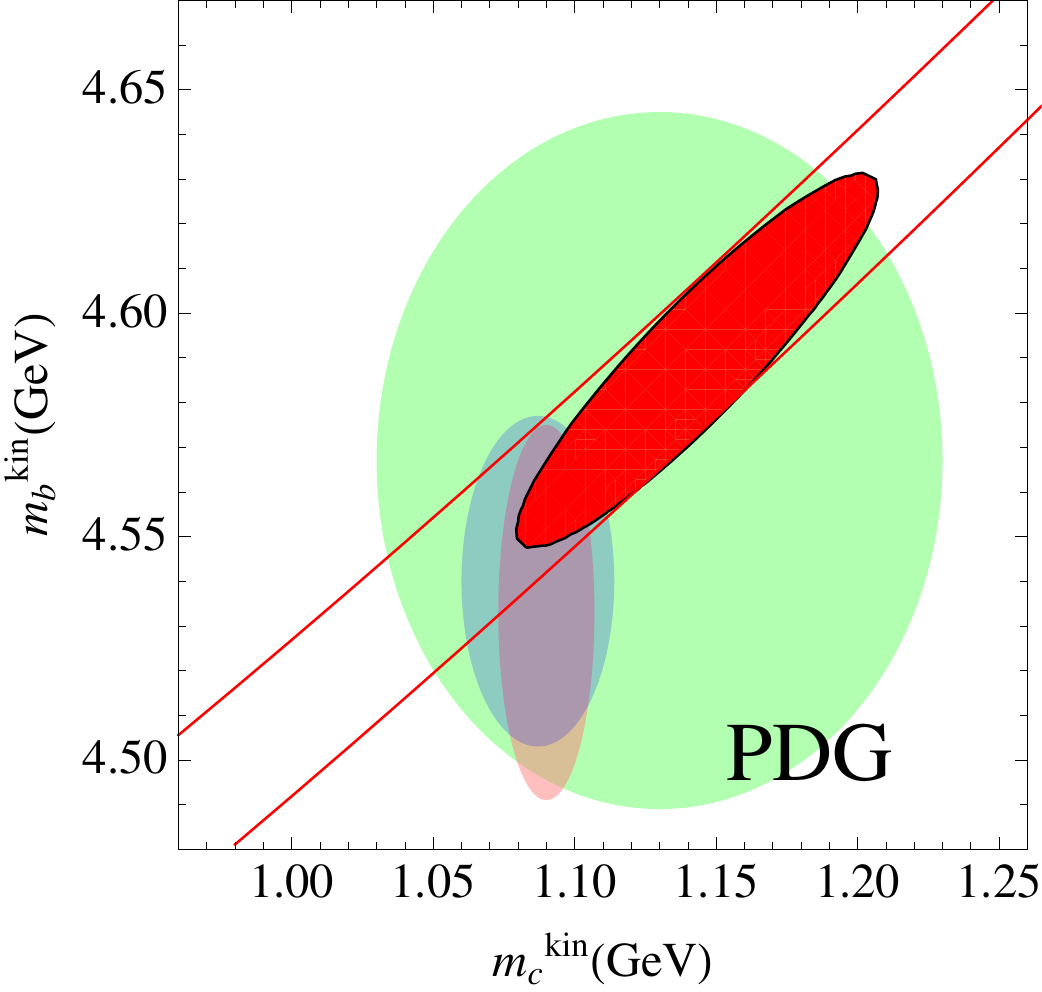}
  \end{center}
  \caption{Different charm and bottom quark determinations in the
  kinetic mass scheme. The ellipses represent the PDG-2007 ranges
  (large green), a global semileptonic fit that differs slightly (see
  text) from the HFAG one (red), the Karlsruhe (pink) and Hoang {\it
  et al.} (blue) sum-rules determinations.} \label{fig:2}
\end{figure}
We have seen that the fits discussed in the previous section determine
$m_b$ quite precisely. How does this $m_b$ determination compare with
alternative determinations~\cite{masses1,hoang,masseslat}?
Semileptonic moments do not measure $m_b$ well. As illustrated in
Fig.~\ref{fig:2}, they rather identify a strip in the $(m_c,m_b)$
plane along which the minimum is quite shallow, and $|V_{cb}|$
basically constant (straight lines). The global kinetic fit selects an
$(m_c,m_b)$ region compatible with the loose PDG-2007~\cite{pdg}
bounds\footnote{Later editions of PDG have stretched the uncertainties
in an abnormal way.} as well as  with the precise $e^+e^-$ sum-rules
determinations~\cite{masses1,hoang}, of course after conversion to the
kinetic scheme. This conversion is known to $O(\alpha_s^2)$ and
entails a non-negligible error, of about 40~MeV for the conversion
from $m_b^{\overline{\rm MS}}(m_b)$ to $m_b^{kin}(1{\rm GeV})$ and
about 10~MeV for that from $m_c^{\overline{\rm MS}}(3{\rm GeV})$ to
$m_c^{kin}(1{\rm GeV})$.

It turns out that the semileptonic fit, and in particular its
determination of the masses and the other OPE parameters,  is very
sensitive to various details. For instance, the assumptions on the
correlations between theoretical errors for moments evaluated at
different cuts have a clear impact on the OPE parameters, while the
value of $|V_{cb}| $ remains quite stable.  Such theoretical
correlations are obviously hard to estimate. The present HFAG fit
follows the procedure outlined in \cite{Buchmuller:2005zv}, assuming
100\% correlation between moments calculated at different values of
$E_{cut}$, the lower cut on the lepton energy. This very strong
assumption distorts the fit, leading to high values of $m_{b,c}$, even
outside the PDG range, and to underestimating  the uncertainty of all
non-perturbative parameters. On the opposite extreme, no correlation
between close values of $E_{cut}$ is unreasonable. A more realistic
approach, adopted in the fit shown in Fig.~2, consists in taking into
account the $E_{cut}$ dependence and correlations of the known OPE
calculation. It leads to slightly lower $m_{b,c}$ with larger
errors. A detailed discussion will be presented elsewhere~\cite{new}.

A related question concerns the role of radiative moments in the fits:
as shown above they help fixing $m_b$. But the fit is almost identical
if one replaces them with the loose bound $m_b^{\overline{\rm
MS}}(m_b)=4.20(7)$~GeV given by PDG in 2007. Indeed, the inclusion of
external, well-founded constraints in the fit can be very useful: it
decreases the errors and neutralizes the potential weight of
theoretical correlations. As semileptonic decays do determine
precisely a linear combination of $m_{b,c}$, a way to maximally
exploit their potential consists in fitting directly
$m_c^{\overline{\rm MS}}(3{\rm GeV})$ instead of the kinetic charm
mass (this is possible and avoids the scheme conversion error), and
including in the fit one of the recent very precise $m_c$
determinations. As an illustration we have used $m_c^{\overline{\rm
MS}}(3{\rm GeV})=0.986(13)$~GeV by the Karlsruhe group~\cite{masses1},
and obtained $m_b^{kin}(1{\rm GeV})=4.535(21)$~GeV, which translates
into $m_b^{\overline{\rm MS}}(m_b)=4.165(45)$~GeV. This value for the
bottom mass is perfectly consistent with the Karlsruhe group's own
$m_b$ determination, $m_b^{\overline{\rm MS}}(m_b)=4.163(16)$~GeV. The
results of Refs.~\cite{hoang} and \cite{masseslat} are also
consistent.

The kinetic scheme fitting routines are now undergoing a major
upgrade, concerning the inclusion of higher order effects, the
possibility to change the perturbative scheme, and the inclusion of
additional constraints in the fit. The preliminary results we have
just shown indicate that an uncertainty of about 20~MeV on $m_b$ can
be reliably reached if an independent, precise determination of $m_c$
is employed. In view of this progress and of the calculations recently
completed or under way, we believe that a  1\% determination of
$|V_{cb}|$ can be reached, although some work is still necessary.

\subsection*{Acknowledgements}
P.G.\ is grateful to N.\ Uraltsev for useful and informative
discussions.


\begin{thebibliography}{99}

\bibitem{Bigi:1992su}
  I.~I.~Y.~Bigi, N.~G.~Uraltsev and A.~I.~Vainshtein,
  Phys.\ Lett.\ B {\bf 293} (1992) 430 [Erratum-ibid.\  B {\bf 297} (1993) 477]
  [arXiv:hep-ph/9207214];
  I.~I.~Y.~Bigi, M.~A.~Shifman, N.~G.~Uraltsev and A.~I.~Vainshtein,
  Phys.\ Rev.\ Lett.\ {\bf 71} (1993) 496 [arXiv:hep-ph/9304225].

\bibitem{Blok:1993va}
  B.~Blok, L.~Koyrakh, M.~A.~Shifman and A.~I.~Vainshtein,
  Phys.\ Rev.\  D {\bf 49} (1994) 3356 [Erratum-ibid.\ D {\bf 50} (1994) 3572]
  [arXiv:hep-ph/9307247];
  A.~V.~Manohar and M.~B.~Wise, Phys.\ Rev.\ D {\bf 49} (1994) 1310 [arXiv:hep-ph/9308246].

\bibitem{Benson:2003kp}
  D.~Benson, I.~I.~Bigi, T.~Mannel and N.~Uraltsev,
  Nucl.\ Phys.\  B {\bf 665}, 367 (2003)
  [hep-ph/0302262].

\bibitem{Gambino:2004qm}
  P.~Gambino and N.~Uraltsev,
  Eur.\ Phys.\ J.\  C {\bf 34}, 181 (2004)
  [hep-ph/0401063].

\bibitem{Benson:2004sg}
  D.~Benson, I.~I.~Bigi and N.~Uraltsev,
  Nucl.\ Phys.\  B {\bf 710}, 371 (2005)
  [hep-ph/0410080].

\bibitem{Bauer:2004ve}
  C.~W.~Bauer, Z.~Ligeti, M.~Luke, A.~V.~Manohar and M.~Trott,
  Phys.\ Rev.\  D {\bf 70}, 094017 (2004)
  [hep-ph/0408002].

 \bibitem{pertb}
  V.~Aquila, P.~Gambino, G.~Ridolfi and N.~Uraltsev,
  Nucl.\ Phys.\  B {\bf 719} (2005) 77
  [arXiv:hep-ph/0503083] and refs.\ therein.

\bibitem{nonpert} M.~Gremm and A.~Kapustin,  {\it Phys.\ Rev.}\ {\bf D55} (1997) 6924.

\bibitem{czarnecki-pak}
  A.~Pak and A.~Czarnecki,
  Phys.\ Rev.\ Lett.\  {\bf 100} (2008) 241807
  [arXiv:0803.0960 [hep-ph]].

\bibitem{melnikov}
  K.~Melnikov,
  Phys.\ Lett.\  B {\bf 666} (2008) 336
  [arXiv:0803.0951 [hep-ph]];
  S.~Biswas and K.~Melnikov,
  JHEP {\bf 1002} (2010) 089
  [arXiv:0911.4142 [hep-ph]].

\bibitem{Becher:2007tk}
  T.~Becher, H.~Boos and E.~Lunghi,
  JHEP {\bf 0712}, 062 (2007)
  [arXiv:0708.0855 [hep-ph]].

\bibitem{ewerth}
  T.~Ewerth, P.~Gambino and S.~Nandi,
  Nucl.\ Phys.\  B {\bf 830}, 278 (2010)
  [arXiv:0911.2175 [hep-ph]].

\bibitem{Mannel:2010wj}
  T.~Mannel, S.~Turczyk and N.~Uraltsev,
  JHEP {\bf 1011}, 109 (2010)
  [arXiv:1009.4622 [hep-ph]].
  
  \bibitem{Bigi:2009ym}
  I.~Bigi {\it et al.},
  JHEP {\bf 1004}, 073 (2010)
  [arXiv:0911.3322 [hep-ph]].
  
  \bibitem{paz} G.~Paz, these proceedings, arXiv:1011.4953 [hep-ph].
  
\bibitem{Aubert:2009qda}
  B.~Aubert {\it et al.}  [BABAR Collaboration],
  Phys.\ Rev.\  D {\bf 81}, 032003 (2010)
  [arXiv:0908.0415 [hep-ex]].

\bibitem{Aubert:2004td}
  B.~Aubert {\it et al.}  [BABAR Collaboration],
  Phys.\ Rev.\  D {\bf 69}, 111104 (2004)
  [arXiv:hep-ex/0403030].

\bibitem{Aubert:2005cua}
  B.~Aubert {\it et al.}  [BABAR Collaboration],
  Phys.\ Rev.\  D {\bf 72}, 052004 (2005)
  [arXiv:hep-ex/0508004].

\bibitem{Aubert:2006gg}
  B.~Aubert {\it et al.}  [BaBar Collaboration],
  Phys.\ Rev.\ Lett.\  {\bf 97}, 171803 (2006)
  [arXiv:hep-ex/0607071].

\bibitem{Urquijo:2006wd}
  P.~Urquijo {\it et al.},
  Phys.\ Rev.\  D {\bf 75}, 032001 (2007)
  [arXiv:hep-ex/0610012].

\bibitem{Schwanda:2006nf}
  C.~Schwanda {\it et al.}  [BELLE Collaboration],
  Phys.\ Rev.\  D {\bf 75}, 032005 (2007)
  [arXiv:hep-ex/0611044].

\bibitem{Hocker:1995kb}
  A.~Hocker and V.~Kartvelishvili,
  Nucl.\ Instrum.\ Meth.\  A {\bf 372}, 469 (1996)
  [arXiv:hep-ph/9509307].

\bibitem{Schwanda:2008kw}
  C.~Schwanda {\it et al.}  [Belle Collaboration],
  Phys.\ Rev.\  D {\bf 78}, 032016 (2008)
  [arXiv:0803.2158 [hep-ex]].


\bibitem{Aubert:2004tea}
  B.~Aubert {\it et al.}  [BABAR Collaboration],
  Phys.\ Rev.\  D {\bf 69}, 111103 (2004)
  [arXiv:hep-ex/0403031].

\bibitem{:2009qg}
  A.~Limosani {\it et al.}  [Belle Collaboration],
  Phys.\ Rev.\ Lett.\  {\bf 103}, 241801 (2009)
  [arXiv:0907.1384 [hep-ex]].

\bibitem{Acosta:2005qh}
  D.~E.~Acosta {\it et al.}  [CDF Collaboration],
  Phys.\ Rev.\  D {\bf 71}, 051103 (2005)
  [arXiv:hep-ex/0502003].

\bibitem{Csorna:2004kp}
  S.~E.~Csorna {\it et al.}  [CLEO Collaboration],
  Phys.\ Rev.\  D {\bf 70}, 032002 (2004)
  [arXiv:hep-ex/0403052].

\bibitem{Chen:2001fja}
  S.~Chen {\it et al.}  [CLEO Collaboration],
  Phys.\ Rev.\ Lett.\  {\bf 87}, 251807 (2001)
  [arXiv:hep-ex/0108032].

\bibitem{Abdallah:2005cx}
  J.~Abdallah {\it et al.}  [DELPHI Collaboration],
  Eur.\ Phys.\ J.\  C {\bf 45}, 35 (2006)
  [arXiv:hep-ex/0510024].

\bibitem{masses1}
  K.~G.~Chetyrkin 
  {\it et al.},
  Phys.\ Rev.\  {\bf D80 } (2009)  074010
  [arXiv:0907.2110 [hep-ph]].

\bibitem{hoang}
  A.~H.~Hoang,
  hep-ph/0008102, and these proceedings.

  \bibitem{masseslat} 
  I.~Allison {\it et al.} [ HPQCD Collaboration ],
  Phys.\ Rev.\  {\bf D78}, 054513 (2008)
  [arXiv:0805.2999 [hep-lat]];
  C.~McNeile {\it et al.} [ HPQCD Collaboration ],
  Phys.\ Rev.\  D {\bf 82} (2010) 034512
  [arXiv:1004.4285 [hep-lat]].

\bibitem{pdg}
W.-M. Yao {\it et al.} (Particle Data Group), J. Phys. G {\bf 33}, 1 (2006) and 2007 partial update for the 2008 edition.

\bibitem{Buchmuller:2005zv}
  O.~Buchmuller, H.~Flacher,
  Phys.\ Rev.\  {\bf D73 } (2006)  073008.
  [hep-ph/0507253].
 
\bibitem{new} P.~Gambino and C.~Schwanda, in progress.

\end{thebibliography}
\end{document}